\documentclass[preprint,preprintnumbers,showpacs,amsmath,amssymb,nofootinbib]{revtex4} 

\usepackage{setspace}

\usepackage{multirow}

\usepackage{float}

\usepackage{graphicx}
\usepackage{epsfig}
\usepackage{amssymb}
\usepackage{graphicx}
\usepackage{dcolumn}
\usepackage{bm}
\newcommand{\ba}{\begin{array}}
\newcommand{\ea}{\end{array}}
\def\br{\begin{eqnarray}}
\def\er{\end{eqnarray}}
\def\be{\begin{equation}}
\def\ee{\end{equation}}

\def\({\left(}
\def\){\right)}

\date{\today}

\begin{document}


\title{Forward Elastic Scattering and Pomeron Models}

\author{M. Broilo$^{1}$\footnote{mateus.broilo@ufrgs.br}, E.G.S. Luna$^{1}$\footnote{luna@if.ufrgs.br} and M.J. Menon$^{2}$\footnote{menon@ifi.unicamp.br}}
\affiliation{$^{1}$Instituto de F\'{\i}sica,  Universidade Federal do Rio Grande do Sul \\ CEP 91501-970, Porto Alegre -- RS, Brazil \\ 
$^{2}$Instituto de F\'{\i}sica Gleb Wataghin, Universidade Estadual de Campinas \\ CEP 13083-859, Campinas -- SP, Brazil}  
 
 \date{\today}

\begin{abstract}
Recent data from LHC13 by the TOTEM Collaboration have indicated an unexpected decrease
in the value of the $\rho$ parameter and a $\sigma_{tot}$ value in agreement with the trend of previous
measurements at 7 and 8 TeV. These data at 13 TeV are not simultaneously described by the predictions from Pomeron models
selected by the COMPETE Collaboration, but show agreement with the maximal Odderon dominance,
as recently demonstrated by Martynov and Nicolescu.
Here we present a detailed analysis on the applicability of Pomeron dominance by means of a general class of
forward scattering amplitude, consisting of even-under-crossing leading contributions
associated with single, double and triple poles in the complex angular momentum plane
and subleading even and odd Regge contributions.
The analytic connection between $\sigma_{tot}$ and $\rho$
is obtained by means of singly subtracted dispersion relations and we carry out fits to $pp$ and
$\bar{p}p$ data in the interval 5 GeV - 13 TeV. The data set comprises all the accelerator data below 7 TeV 
and we consider
two independent ensembles by adding either only the TOTEM data or TOTEM and ATLAS data at
the LHC energy region. In the data reductions to each ensemble the uncertainty regions are
evaluated with both one and two standard deviation ($\sim$ 68 \% and $\sim$ 95 \% CL, respectively).
Besides the general analytic model,
we investigate four particular cases of interest, three of them typical of outstanding models in the        
literature. 
We conclude that, within the experimental and theoretical uncertainties and both
ensembles, the general model and three particular cases are not able to describe
the $\sigma_{tot}$ and $\rho$ data at 13 TeV simultaneously.
However, if the discrepancies between the TOTEM and ATLAS data are not resolved,
one Pomeron model, associated with double and triple poles and with only 7 free parameters, seems not to be 
excluded by the complete set of experimental information presently available.
\end{abstract}


\maketitle

\section{Introduction}

In elastic hadron-hadron collisions, the forward scattering is characterized
by two physical observables, the total cross section, $\sigma_{tot}$, and
the $\rho$ parameter. In terms of the scattering amplitude $\mathcal{A}$ and its
Mandelstam variables ($s$ and $t$, energy and momentum transfer squared in the c.m. system),
the former is given by the optical
theorem, which at high energies can be expressed by \cite{pred}
\begin{eqnarray}
\sigma_{\mathrm{tot}}(s) = \frac{\mathrm{Im}\,\mathcal{A}(s,t=0)}{s},
\label{st}
\end{eqnarray}
and the later, associated with the phase of the amplitude, is defined by
\begin{eqnarray}
\rho(s) = \frac{\mathrm{Re}\,\mathcal{A}(s,t=0)}{\mathrm{Im}\,\mathcal{A}(s,t=0)},
\label{rho}
\end{eqnarray}
where $t=0$ indicates the forward direction.

Since the real and imaginary parts of the amplitude can be formally correlated
by means of dispersion relations, Eqs. (\ref{st}) and (\ref{rho}) provide a fundamental
physical connection between the phase of the amplitude ($\rho$) and the total probability
of the hadronic interaction ($\sigma_{tot}$), as a function of the energy.
However, despite their rather simple analytic forms, the investigation of these two quantities
in terms of the energy, constitute a long-standing challenge in the study of the
hadronic interactions \cite{giulia}. 

In the experimental context, to access the forward and near forward region demands complex and
sophisticated instrumentation and data analyses. In addition, the difficulties grow progressively as the
energy increases. For example, the $\rho$ parameter is determined in the region of
interference between the Coulomb and hadronic interactions, which are of the same magnitude
at values of the momentum transfer proportional to the inverse of the total cross section
(see for example \cite{block}, Sect. 4). As a consequence of the rise of $\sigma_{tot}$
at the highest energies, it becomes extremely difficult to reach this region as the
energy increases.

In the theoretical QCD context,
this deep (extreme) infrared region ($t \rightarrow 0$) is not expected to be accessed by perturbative techniques.
A crucial point concerns the absence of a nonperturbative approach
able to predict the energy dependence of $\sigma_{tot}$
and $\rho$ from first principles and without model assumptions.

In the phenomenological context, beyond classes of models including other physical 
quantities\footnote{For recent reviews see, for example, \cite{giulia,block,dremin}.},
$\sigma_{tot}(s)$ and $\rho(s)$ are usually investigated by means of
\textit{amplitude analyses}, an approach based on the Regge-Gribov theory and analytic $S$-Matrix concepts.
In this formalism \cite{collins,fr}, the singularities in the complex angular momentum
$J$-plane ($t$-channel) are associated with the asymptotic behavior of the elastic scattering
amplitude in terms of the energy ($s$-channel). 
In the general case, associated with a pole of order $N$, the contribution to the 
imaginary part of the \textit{forward} amplitude
in the $s$-channel
is $s^{\alpha_0} \ln^{N-1}(s)$, where $\alpha_0$ is the intercept of the trajectory
(see Appendix B in \cite{fms17b} for a recent short review).
Therefore, for the total cross section we have 
\begin{eqnarray}
\sigma_{\mathrm{tot}}(s)  \propto s^{\alpha_0 - 1}\ln^{N-1} s,
\nonumber
\end{eqnarray}
and the following possibilities connecting the singularities ($J$-plane) and the asymptotic behavior:

\begin{itemize}

\item
simple pole ($N=1$) at $J = \alpha_0$, with $\alpha_0 = 1$\ \ $\Rightarrow$\ \ $\sigma$ \ constant;

\item
simple pole ($N=1$) at $J=\alpha_0$\ \ $\Rightarrow$\ \ $\sigma \propto s^{\alpha_0-1}$;

\item
double pole ($N=2$) at $J=\alpha_0$, with $\alpha_0=1$\ \ $\Rightarrow$\ \ $\sigma \propto \ln(s)$;

\item
triple pole ($N=3$) at $J=\alpha_0$, with $\alpha_0=1$\ \ $\Rightarrow$\ \ $\sigma \propto \ln^2(s)$.

\end{itemize}

For an elastic particle-particle and antiparticle-particle scattering, given the above inputs
for $\sigma_{tot}(s)$, through Eq. (1), the even and odd contributions associated with
$\mathrm{Im}\,\mathcal{A}(s,t=0) / s$ are defined (Crossing) and the corresponding real parts
are obtained by means of dispersion relations (Analyticity), leading to $\rho(s)$
in Eq. (2).

Historically, the leading contribution to $\sigma_{tot}$ at the highest energies
has been associated with an even-under-crossing object named Pomeron (from a QCD viewpoint, a color singlet
made up of two gluons in the simplest configuration) \cite{fr}. 
Typical Pomeron models consider contributions associated
with either a simple pole at $J=\alpha_0$ (for example, Donnachie and Landshoff \cite{dl79}
and some QCD-inspired models \cite{bghp}) or
a triple pole at $J=1$ (as selected in the detailed analysis by the COMPETE Collaboration \cite{compete1,compete2}
and used in the successive editions of the Review of Particle Physics, by the
Particle Data Group (PDG) \cite{pdg16}).

However, recently, new experimental information on $\sigma_{\mathrm{tot}}$ and $\rho$ from LHC13 
were reported by the TOTEM Collaboration \cite{totem1,totem2}:
\begin{eqnarray}
\sigma_{\mathrm{tot}} &=& 110.6 \pm 3.4\ \mathrm{mb}, \nonumber \\
\rho &=& 0.10 \pm 0.01\ \mathrm{and}\ \rho = 0.09 \pm 0.01,
\label{totem13}
\end{eqnarray}
indicating an unexpected decrease in the value of the $\rho$ parameter and $\sigma_{tot}$ in agreement
with the trend of previous measurements at 7 and 8 TeV.
Indeed, recent investigation concerning bounds on the rise of $\sigma_{tot}(s)$, including all
TOTEM data at 7 and 8 TeV and the L$\gamma$ parameterization \cite{fms17a}, has predicated at 13 TeV
the value $\sigma_{tot} =$ 110.7 $\pm$ 1.2 mb, which is in full agreement with the above measurement.
However, for $\rho$ at 13 TeV the extrapolation yielded 0.1417 $\pm$ 0.0047, indicating complete disagreement 
with the data and
far above the experimental result (see Table 4 in \cite{fms17a}).
Moreover, the results (\ref{totem13}) are not simultaneously described
by all the predictions of the Pomeron models from the detailed analysis by the COMPETE
Collaboration in 2002 \cite{compete1}, as pedagogically shown in Figure 18 of \cite{totem2}). 

Remarkably,
the odd-under-crossing asymptotic contribution,
introduced by  Lukaszuk and Nicolescu \cite{odderon} and
named Odderon \cite{oddname} (from a QCD viewpoint, a color-singlet made up of three gluons in the simplest configuration) \cite{braun},
provides quite good descriptions of the
experimental data, as predicted by the Avila-Gauron-Nicolescu model \cite{agn} and
demonstrated recently in the analyses by Martynov and Nicolescu
\cite{martynico1,martynico2}.

On the other hand, also recently, the above data at 13 TeV have been analyzed by
Khoze, Martin and Ryskin
in the context of a QCD-based multichannel eikonal model (Pomeron dominance),
tuned in 2013 with data up to 7 TeV \cite{kmr14}. The analysis indicates that the
data at 13 TeV are reasonably described without an odd-signature term \cite{kmr1}.
Moreover, the authors also argument that the 
maximal Odderon is inconsistent with the black-disk limit \cite{kmr2}.
Very recently, subsequent articles have also discussed possible effects related to Odderon
contributions in different contexts \cite{kmr}.

In view of all these recent information and the fact that \textit{forward amplitude analyses}
have favored the Pomeron dominance, at least, up to 8 TeV, it seems important to develop detailed tests on the
applicability of the Pomeron models by means of a general class of forward scattering amplitudes.
With that in mind, we have already reported two forward analyses with
Pomeron dominance and including, for the first time, the TOTEM data at 13 TeV.
In the first work, two models have been tested, without taking into account
the uncertainty regions in the data reductions \cite{blm18a} (as in the Martynov and Nicolescu analyses
\cite{martynico1,martynico2}). We concluded that the models are not able to
satisfactorily describe the $\sigma_{tot}$ and $\rho$ data at 13 TeV.
In the second analysis, we have considered one Pomeron model with 6 free parameters
and have evaluated the uncertainty regions with confidence level (CL) of
90 \%. We have concluded that the model seems not to be excluded by the
bulk of experimental data presently available \cite{blm18b}.

In this paper, we shall extend our investigation in several important aspects.
The main point is to consider 
\textit{classes of even leading contributions} by incorporating 
different components of several models and investigating the effect of several
combinations, with focus on the \textit{uncertainties involved in the data reductions}. 
To this end, we shall treat a general parameterization
for $\sigma_{\mathrm{tot}}(s)$ consisting of constant, power, logarithmic and logarithmic
squared functions of the energy, together with
even and odd Reggeons ($a_2/f_2$ and $\rho/\omega$ trajectories, respectively)
for the low energy region.
The analytic connection with  $\rho(s)$ is obtained by means of even and odd
singly subtracted dispersion relations.
We carry out fits to $pp$ and $\bar{p}p$ data on $\sigma_{tot}$ and $\rho$
in the interval 5 GeV - 13 TeV through the general parameterization as well
as \textit{four particular cases}, three of them typical of outstanding models in the
literature.

However, there is an intrinsic difficulty with this kind of analysis 
deserving attention from the beginning. In what concerns the $\sigma_{tot}$ and $\rho$
data, despite the great expectations with the LHC, the experimental information
presently available are characterized by discrepancies between the measurements 
of $\sigma_{tot}$ by the TOTEM Collaboration and
by the ATLAS Collaboration  at 7 TeV and mainly at 8 TeV. Some consequences
of these discrepancies have already been discussed
by Fagundes, Menon and Silva \cite{fms17a} and by us in \cite{blm18b}.
In their first analysis, Martynov and Nicolescu present arguments for not including the ATLAS data 
\cite{martynico1}, which, however, have been included in their second analysis \cite{martynico2}.
We shall return to these important topics along the paper.

It should be also noted that the uncertainties in the TOTEM measurements of $\sigma_{tot}$ are essentially systematic
(uniform distribution) and not statistical (Gaussian distribution). This fact puts limitations in a strict interpretation
of the $\chi^{2}$ test of goodness-of-fit. This point is discussed in certain detail in Ref. \cite{fms17b}, Table 1 and
Appendix A, specially Sect. A.1. 

Here, to address the above question and as in previous analyses \cite{fms17a,fms17b,blm18a,blm18b},
we shall adopt two variants for defining our data set:
all the experimental data below 7 TeV (above 5 GeV) and two independent
ensembles by adding either only the TOTEM data at 7, 8 and 13 TeV (ensemble T) or by including also the
ATLAS data at 7 and 8 TeV (ensemble T+A).
In addition, in order to investigate and stress the importance of the uncertainty regions
in the fit results,
we shall consider data reductions with two different CL, associated with both one and two
standard deviations ($\sigma$), namely 68.27 \% and 95.45 \% CL respectively.

Taking into account the aforementioned critical remarks related to the LHC data,
as well as, within
the theoretical and experimental uncertainties and both ensembles, our main 
conclusions from the data 
reductions are the following:
(1) the general analytic model and three particular cases cannot describe,
simultaneously, the $\sigma_{tot}$ and $\rho$ data at 13 TeV;
(2) one particular Pomeron model, with only 7 free parameters and associated
with double and triple poles, seems not to be excluded
by the bulk of experimental data presently available;
(3) for this Pomeron model additional tests on the effect of the subtraction
constant and the energy cutoff for data reductions, select the constant as
a free fit parameter and cutoff at 5 GeV.
These results corroborate our previous conclusion \cite{blm18a}, now
with 1 $\sigma$ and 2 $\sigma$. 

The manuscript is organized as follows. The analytic models
are introduced in Sect. 2 and the fit procedures and results
are presented in Sect. 3. In Sect. 4 we discuss all the results
and in Sect. 5 we present our conclusions and final remarks.
In an Appendix it is treated some additional tests together with
discussions on the results.

\section{Analytic Models}

As commented in our introduction, in the Regge-Gribov theory, simple, double and triple poles 
in the complex angular momentum
plane are associated with power, logarithmic and logarithmic squared functions 
of the energy for the total
cross section.
In this context, for $pp$ and $\bar{p}p$ scattering,  we consider a general parameterization for $\sigma_{tot}(s)$
consisting of two Reggeons (even and odd under crossing) and four (even) Pomeron
contributions:

\begin{eqnarray}
\sigma_{\mathrm{tot}}(s) =  a_1 \left[\frac{s}{s_0}\right]^{-b_1} \!\!\!\!\!\!+ \tau a_2 \left[\frac{s}{s_0}\right]^{-b_2}
\!\!\!\!\!\!+ A + B \left[\frac{s}{s_0}\right]^{\epsilon}+ C \ln\left(\frac{s}{s_0}\right) + D \ln^2\left(\frac{s}{s_0}\right),
\label{stg}
\end{eqnarray}
where $\tau = -1$ for $pp$, $\tau = +1$ for $\bar{p}p$, while
$a_1$, $b_1$, $a_2$, $b_2$, are free fit parameters associated with the secondary Reggeons,
$A$, $B$, $\epsilon$, $C$, $D$ are the free parameters associated with Pomeron components and $s_0$ is an
energy scale to be discussed in what follows.

The analytic results for $\rho(s)$ have been obtained by means of singly subtracted 
derivative dispersion relations \cite{bks}, taking into account an effective subtraction 
constant $K$:
\begin{eqnarray}
\!\!\!\!\!\!\rho(s) &=& \frac{1}{\sigma_{\mathrm{tot}}(s)}
\left\{ \frac{K}{s}  
- a_1\,\tan \left( \frac{\pi\, b_1}{2}\right) \left[\frac{s}{s_0}\right]^{-b_1} \!\!\!\!\!\!+
\tau \, a_2\, \cot \left(\frac{\pi\, b_2}{2}\right) \left[\frac{s}{s_0}\right]^{-b_2} \right. \nonumber \\
&+& \left. B\,\tan \left( \frac{\pi\, \epsilon}{2}\right) \left[\frac{s}{s_0}\right]^{\epsilon} +
\frac{\pi C}{2} + \pi D \ln\left(\frac{s}{s_0}\right) \right\}.
\label{rhog}
\end{eqnarray}
As discussed in detail in Appendix C of \cite{fms17b} (and quoted references), $K$
avoids
the full high-energy approximation in dispersion relation approaches. 

Here, following \cite{fms17b,fms17a,blm18a,blm18b}, we consider the energy scale \textit{fixed} at
the physical threshold for scattering states, 
\begin{eqnarray}
s_0 = 4m_p^2 \sim  3.521 \mathrm{GeV}^2,
\nonumber
\end{eqnarray}
where $m_p$ is the proton mass (see Sect. 4.3 in \cite{ms13b} for discussions on this choice).

Eqs. (\ref{stg}) and (\ref{rhog}) bring enclosed analytic structures \textit{similar} to those 
appearing in some well known models,
as for example, Donnachie and Landshoff (DL) \cite{dl79}, Block and Halzen (BH) \cite{bh1},
COMPETE and PDG parameterizations (COMPETE) \cite{compete1,compete2,pdg16}. We shall consider four particular cases,
distinguished by the corresponding Pomeron contributions ($\sigma^P$), defined and denoted as follows.

\begin{itemize}

\item Model I:\
$A=C=D=0$ \quad $\Rightarrow$ \quad $\sigma_{I}^{P} = B \left[\frac{s}{s_0}\right]^{\epsilon}$ \quad (DL-type)

\item Model II:\ 
$B=C=0$, $\epsilon=0$ \quad $\Rightarrow$ \quad $\sigma_{II}^{P} = A + D \ln^{2}\left(\frac{s}{s_0}\right)$ \quad  (COMPETE-type)

\item Model III:\
$A=B=0$, $\epsilon=0$ \quad $\Rightarrow$ \quad $\sigma_{III}^{P} = C \ln \left(\frac{s}{s_0}\right) + 
D \ln^{2}\left(\frac{s}{s_0}\right)$ \quad  (BH-type)

\item Model IV:\
$A=D=0$ \quad $\Rightarrow$ \quad $\sigma_{IV}^{P} = B \left[\frac{s}{s_0}\right]^{\epsilon}
+ C \ln \left(\frac{s}{s_0}\right)$
\quad (hybrid power-log)

\end{itemize}

We note that models II and III are analytically similar. The difference concerns the
phenomenological interpretation of the singularities as single and double poles.
Also, in the BH analyses the energy scale is fixed (as we consider here) and in the
COMPETE case it is treated as a free fit parameter.

As far as we know, model IV was never considered in the literature.
Its use here is related to tests on the power law (single pole) in the attempt
to describe simultaneously the $\sigma_{tot}$ and
the $\rho$ data at 13 TeV. We shall return to this point in Sect. IV.B.
 
In the General Model,  Eqs. (\ref{stg}) and (\ref{rhog}), we have 10 free parameters, $a_1$, $b_1$, $a_2$, $b_2$, $A$, $B$, $\epsilon$, $C$, $D$ and $K$, which are determined through fits to the experimental data on $\sigma_{tot}$ and $\rho$ from $pp$ and $\bar{p}p$ elastic scattering in the interval 5 GeV - 13 TeV.

\section{Fits and Results}

\subsection{Ensembles and Data Reductions}

The data above 5 GeV and below 7 TeV have been collected from the PDG database \cite{pdg16}, without
any kind of data selection or sieve procedure (we have used all the
published data by the experimental collaborations). The data at 7 and 8 TeV
by the TOTEM and ATLAS Collaborations can be found in \cite{fms17b}, Table
1, together with further information and complete list of references. 
The TOTEM data at 13 TeV are those in (\ref{totem13}) \cite{totem1,totem2}.

As commented in our introduction, given the tension between the TOTEM and ATLAS measurements on $\sigma_{tot}$
at 7 TeV and mainly 8 TeV, we shall consider two ensembles of
$pp$ and $\bar{p}p$ data above 5 GeV, both comprising the same dataset
in the region below 7 TeV. We then construct:

\begin{itemize}

\item
Ensemble TOTEM (T) by adding only the TOTEM data at 7, 8 and 13 TeV;

\item
Ensemble TOTEM + ATLAS (T + A)  by adding to ensemble T the ATLAS data at 7 and 8 TeV.

\end{itemize}

The fits were performed with the objects of the TMinuit package and using the default MINUIT 
error analysis \cite{minuit}.
We have carried out global fits using a $\chi^{2}$ fitting procedure,
where the value of $\chi^{2}_{min}$ is distributed as a $\chi^{2}$ distribution with $\nu$ degrees of freedom. 
The global fits to $\sigma_{tot}$ and $\rho$ data were performed adopting an interval $\chi^{2}-\chi^{2}_{min}$ corresponding, in the case of normal errors, to the projection of the $\chi^{2}$ hypersurface containing first $\sim$ 68 \% of probability, 
and in a second step, $\sim$ 95 \% of probability,
namely 1 $\sigma$ and 2 $\sigma$.

As a convergence criteria we consider only minimization results which imply positive-definite covariance matrices, since theoretically the covariance matrix for a physically motivated function must be positive-definite at the minimum. 
As tests of goodness-of-fit we shall adopt the chi-square per degree of freedom
$\chi^2/\nu$ and the integrated probability $P(\chi^2)$ \cite{bev}.

\subsection{Fit Results}

The data reductions with the general model given by Eqs. (\ref{stg}) and (\ref{rhog}) did not comply with the above convergence requirements and thus can not be regarded as a possible solution. This may be due to an excessive number of free parameters. 
On the other hand,  in the particular cases given by Models I, II, III and IV, the convergence criteria were reached. 

In each case, the values of the free fit parameters with uncertainty of 1 $\sigma$, together with the corresponding statistical information, 
are displayed in Table I in case of ensemble T and Table II within ensemble T+A. 

Through  error propagation from the fit parameters, we determine the uncertainty
regions for the theoretical results (curves), within 1 $\sigma$ and 2 $\sigma$.
The results for $\sigma_{tot}(s)$ and $\rho(s)$ with models I, II, III and IV (ensembles T and T+A) are compared with the experimental data in Figures 1, 2, 3 and 4, respectively. In each Figure, the insets highlight the LHC energy region.

\begin{table*}[!ht]
\caption{Fit results to $\sigma_{tot}$ and $\rho$ data from ensemble T through models I - IV (Sect. II),
by considering one standard deviation, energy cutoff at 5 GeV and $K$ as a free fit parameter.}
\begin{tabular}{c@{\quad}c@{\quad}c@{\quad}c@{\quad}c@{\quad}}
\hline \hline
& & &  \\[-0.4cm]
Model:     & I  & II  & III  & IV    \\[0.05ex]
\hline
& & & &\\[-0.2cm]
$a_{1}$ (mb)        & 41.4\,$\pm$\,1.8       & 32.2\,$\pm$\,1.8       & 58.8\,$\pm$\,1.5       & 51.5\,$\pm$\,7.1\\[0.05ex]
$b_{1}$             & 0.378\,$\pm$\,0.028    & 0.392\,$\pm$\,0.049    & 0.229\,$\pm$\,0.017    & 0.296\,$\pm$\,0.037\\[0.05ex]
$a_{2}$ (mb)        & 17.0\,$\pm$\,2.0       & 17.0\,$\pm$\,2.1       & 16.9\,$\pm$\,2.0       & 17.0\,$\pm$\,2.1\\[0.05ex]
$b_{2}$             & 0.545\,$\pm$\,0.037    & 0.545\,$\pm$\,0.037    & 0.543\,$\pm$\,0.036    & 0.544\,$\pm$\,0.037\\[0.05ex]
$A$ (mb)            & -                      & 29.6\,$\pm$\,1.2       & -                      & - \\[0.05ex]
$B$ (mb)            & 21.62\,$\pm$\,0.73     & -                      & -                      & 9.6\,$\pm$\,7.5\\[0.05ex]
$\epsilon$          & 0.0914\,$\pm$\,0.0030  & -                      & -                      & 0.108\,$\pm$\,0.019\\[0.05ex]
$C$ (mb)            & -                      & -                      & 3.67\,$\pm$\,0.34      & 2.4\,$\pm$\,1.6\\[0.05ex]
$D$ (mb)            & -                      & 0.251\,$\pm$\,0.010    & 0.132\,$\pm$\,0.024    & - \\[0.05ex]
$K$ (mbGeV$^{2}$)   & 69\,$\pm$\,47          & 55\,$\pm$\,50          & 20\,$\pm$\,44          & 45\,$\pm$\,47\\[0.05ex]
\hline
& & & & \\[-0.4cm]
$\nu$               & 248                    & 248                    & 248                    & 247\\[0.05ex]
$\chi^2/\nu$        & 1.273                  & 1.193                  & 1.210                  & 1.249\\[0.05ex]
$P(\chi^2)$         & 2.3 $\times$ 10$^{-3}$ & 2.0 $\times$ 10$^{-2}$ & 1.4 $\times$ 10$^{-2}$ & 4.8 $\times$ 10$^{-3}$\\[0.05ex]
\hline
& & & & \\[-0.4cm]
Figure:             &           1            &           2            &            3           &         4            \\[0.05ex]
\hline \hline 
\end{tabular}
\label{t1}
\end{table*}

\begin{table*}[!ht]
\caption{Fit results to $\sigma_{tot}$ and $\rho$ data from ensemble T+A through models I - IV (Sect. II),
by considering one standard deviation, energy cutoff at 5 GeV and $K$ as a free fit parameter.}
\begin{tabular}{c@{\quad}c@{\quad}c@{\quad}c@{\quad}c@{\quad}}
\hline \hline
& & &  \\[-0.4cm]
Model:     & I  & II  & III  & IV    \\[0.05ex]
\hline
& & & &\\[-0.2cm]
$a_{1}$ (mb)        & 41.4\,$\pm$\,1.8       & 32.3\,$\pm$\,2.0       & 59.1\,$\pm$\,1.5        & 53.1\,$\pm$\,9.6   \\[0.05ex]
$b_{1}$             & 0.386\,$\pm$\,0.028    & 0.412\,$\pm$\,0.045    & 0.234\,$\pm$\,0.016     & 0.291\,$\pm$\,0.044\\[0.05ex]
$a_{2}$ (mb)        & 17.0\,$\pm$\,2.1       & 17.0\,$\pm$\,2.0       & 16.9\,$\pm$\,2.0        & 17.0\,$\pm$\,2.1   \\[0.05ex]
$b_{2}$             & 0.545\,$\pm$\,0.037    & 0.545\,$\pm$\,0.036    & 0.543\,$\pm$\,0.036     & 0.544\,$\pm$\,0.038\\[0.05ex]
$A$ (mb)            & -                      & 30.20\,$\pm$\,0.90     & -                       & -                  \\[0.05ex]
$B$ (mb)            & 22.01\,$\pm$\,0.64     & -                      & -                       & 8.0\,$\pm$\,10   \\[0.05ex]
$\epsilon$          & 0.0895\,$\pm$\,0.0024  & -                      & -                       & 0.110\,$\pm$\,0.033\\[0.05ex]
$C$ (mb)            & -                      & -                      & 3.81\,$\pm$\,0.30       & 2.8\,$\pm$\,2.1    \\[0.05ex]
$D$ (mb)            & -                      & 0.2438\,$\pm$\,0.0077    & 0.119\,$\pm$\,0.020     & -                  \\[0.05ex]
$K$ (mbGeV$^{2}$)   & 73\,$\pm$\,48          & 64\,$\pm$\,50          & 23\,$\pm$\,43           & 46\,$\pm$\,48      \\[0.05ex]
\hline
& & & & \\[-0.4cm]
$\nu$               & 250                    & 250                    & 250                     & 249\\[0.05ex]
$\chi^2/\nu$        & 1.307                  & 1.227                  & 1.234                   & 1.273\\[0.05ex]
$P(\chi^2)$         & 7.9 $\times$ 10$^{-4}$ & 8.2 $\times$ 10$^{-3}$ & 6.9 $\times$ 10$^{-3}$  & 2.3 $\times$ 10$^{-3}$\\[0.05ex]
\hline
& & & & \\[-0.4cm]
Figure:             &           1            &           2            &            3           &         4            \\[0.05ex]
\hline \hline 
\end{tabular}
\label{t2}
\end{table*}

\begin{figure}[H] 
\begin{center}
 \includegraphics[width=8.0cm,height=9.5cm]{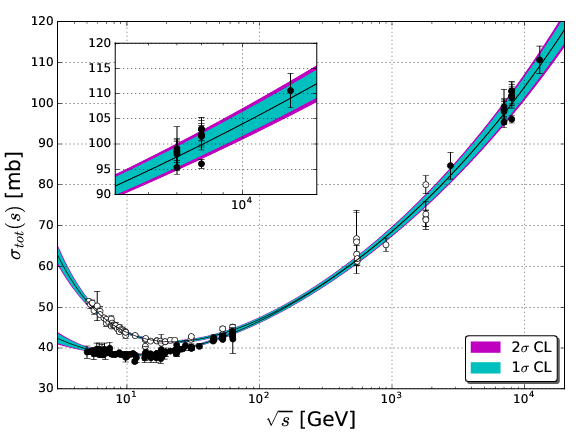}
 \includegraphics[width=8.0cm,height=9.5cm]{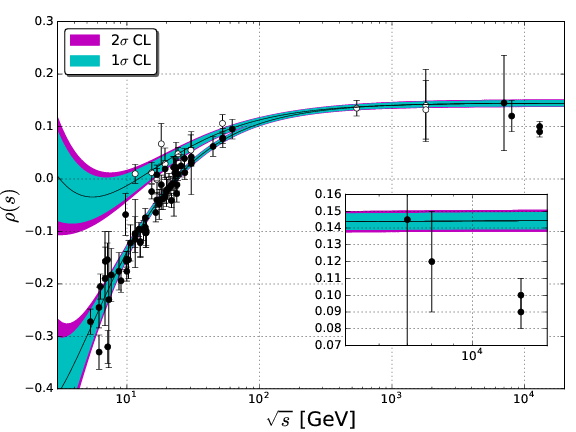}
 \includegraphics[width=8.0cm,height=9.5cm]{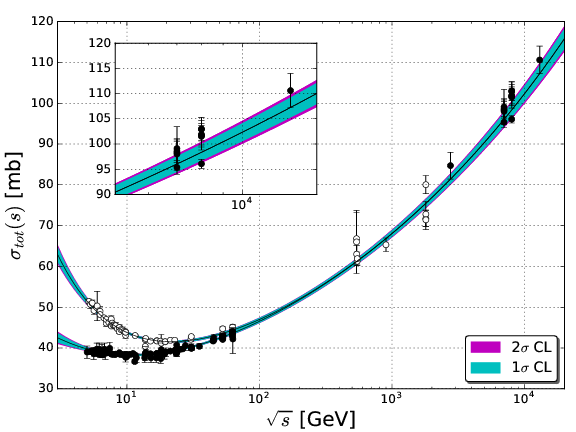}
 \includegraphics[width=8.0cm,height=9.5cm]{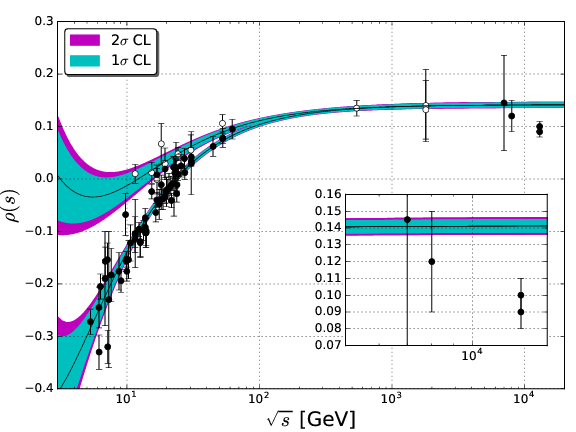}
 \caption{Fit results with Model I to ensembles T (above) and T+A (below).}
\label{f1}
\end{center}
\end{figure}

\begin{figure}[H]
\begin{center}
 \includegraphics[width=8.0cm,height=9.5cm]{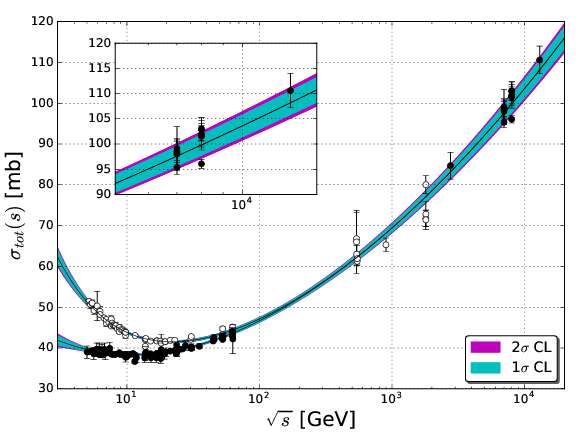}
 \includegraphics[width=8.0cm,height=9.5cm]{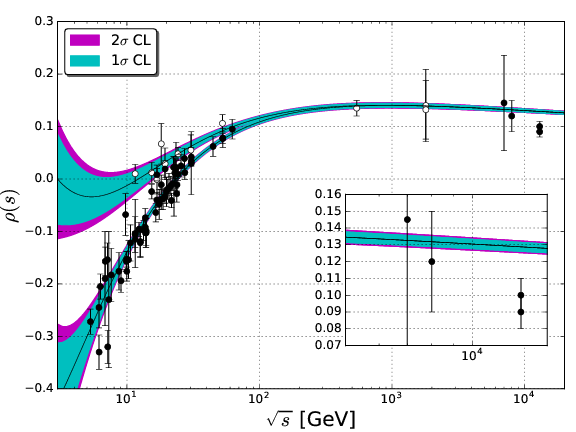}
 \includegraphics[width=8.0cm,height=9.5cm]{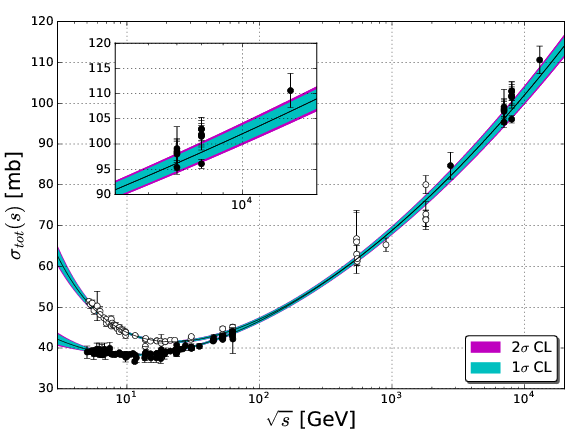}
 \includegraphics[width=8.0cm,height=9.5cm]{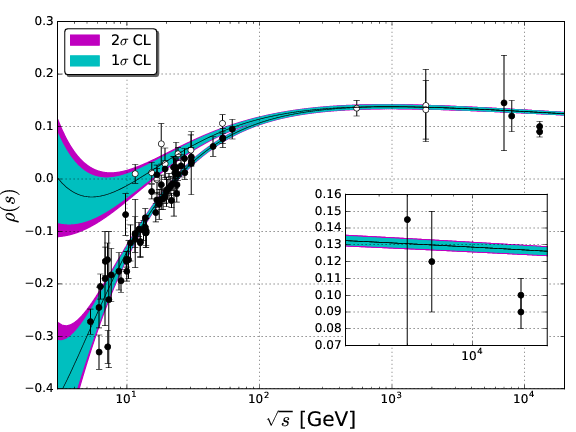}
 \caption{Fit results with Model II to ensembles T (above) and T+A (below).}
\label{f2}
\end{center}
\end{figure}

\begin{figure}[H]
\begin{center}
 \includegraphics[width=8.0cm,height=9.5cm]{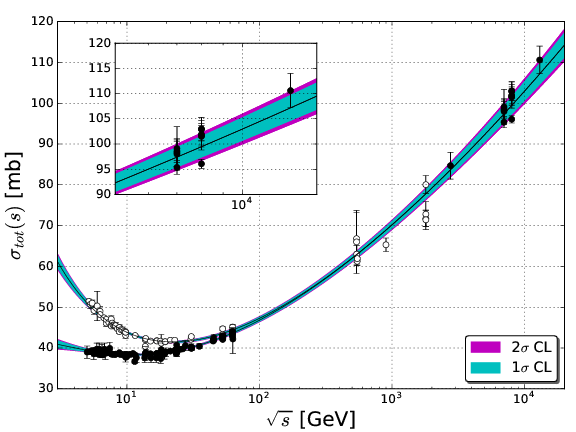}
 \includegraphics[width=8.0cm,height=9.5cm]{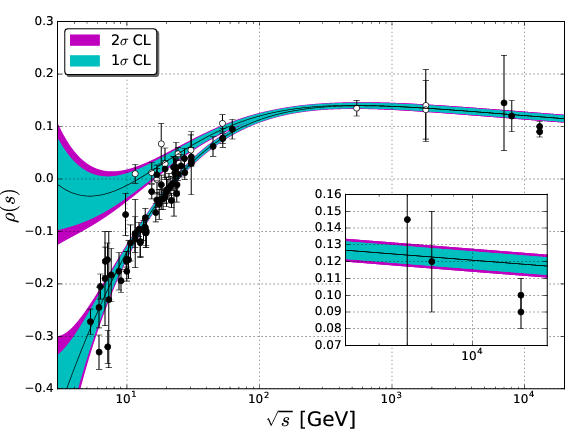}
 \includegraphics[width=8.0cm,height=9.5cm]{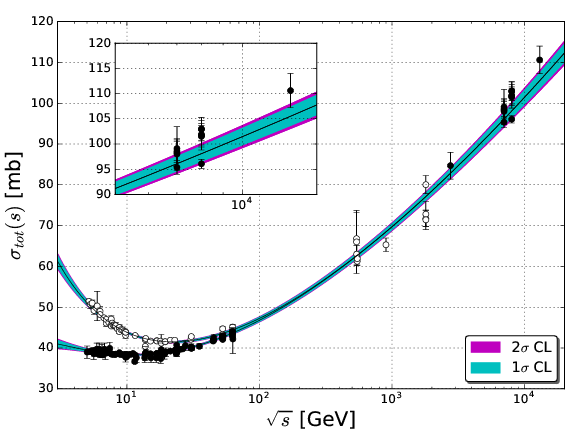}
 \includegraphics[width=8.0cm,height=9.5cm]{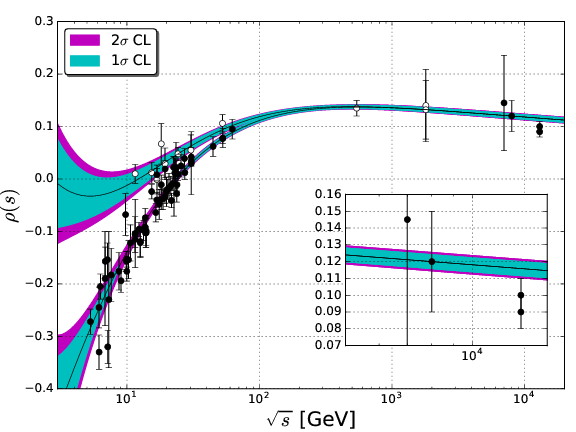}
 \caption{Fit results with Model III to ensembles T (above) and T+A (below).}
\label{f3}
\end{center}
\end{figure}

\begin{figure}[H]
\begin{center}
 \includegraphics[width=8.0cm,height=9.5cm]{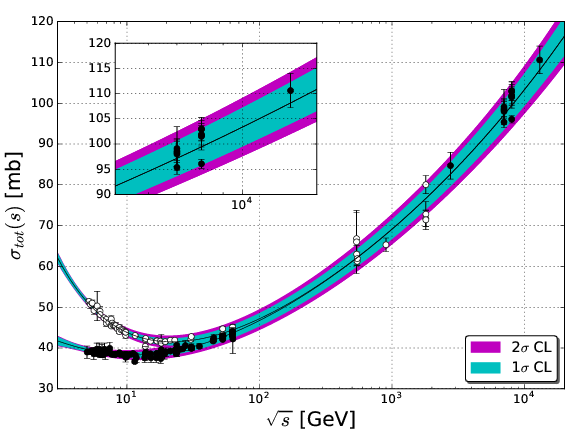}
 \includegraphics[width=8.0cm,height=9.5cm]{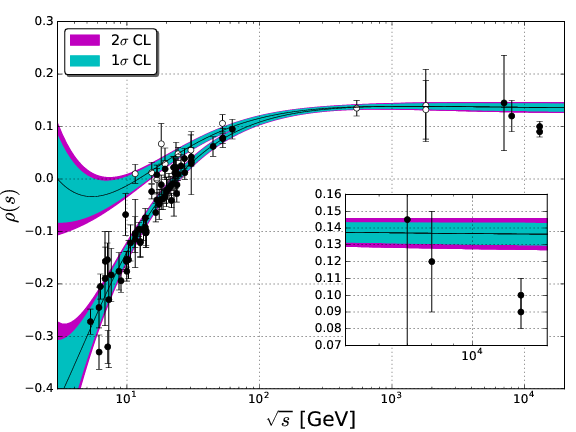}
 \includegraphics[width=8.0cm,height=9.5cm]{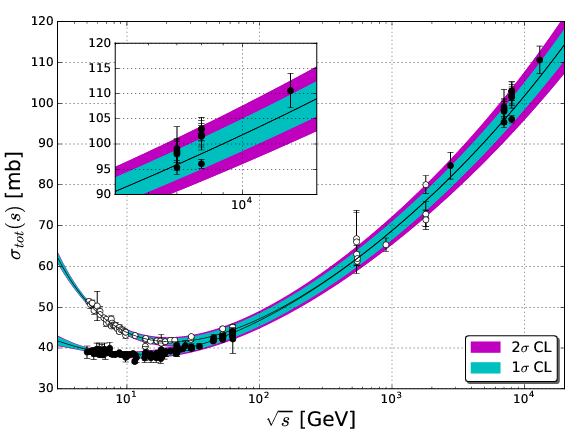}
 \includegraphics[width=8.0cm,height=9.5cm]{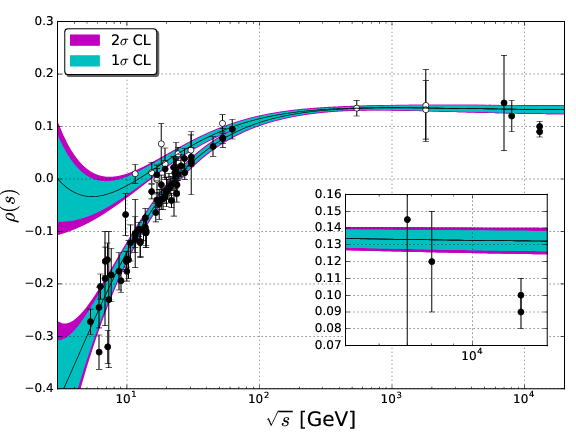}
 \caption{Fit results with Model IV to ensembles T (above) and T+A (below).}
\label{f4}
\end{center}
\end{figure}

\newpage

\section{Discussion}

On the basis of the fit results in Tables I and II and Figures 1-4, let us first separate our 
discussion in two topics related to ensembles (Sub-sec. A) and models (Sub-sec. B). After that
we shall discuss a selected model in more detail (Sub-sec. C).

\subsection{Ensembles T and T+A}

Ensemble T+A encompasses all the experimental data presently available
on forward $pp$ and $\bar{p}p$ scattering at high energies. However, as commented in our introduction
and discussed in \cite{fms17a} and \cite{blm18b},
the TOTEM and ATLAS data at 7 and 8 TeV present discrepant values.
In special, at 8 TeV, the ATLAS measurement of $\sigma_{tot}$ differs from the
latest TOTEM result at this energy by 3 standard deviation,
\begin{eqnarray} 
\frac{\sigma_{tot}^{\mathrm{TOTEM}} - \sigma_{tot}^{\mathrm{ATLAS}}}{\Delta \sigma_{tot}^{\mathrm{TOTEM}}} = 
\frac{103 - 96.07}{2.3} = 3.0.
\nonumber
\end{eqnarray} 

On the one hand, TOTEM published 4 measurements at 7 TeV and 5 at 8 TeV
(all consistent among them) and ATLAS only one point at each energy.
On the other hand, the ATLAS uncertainties in these results are much smaller then
the TOTEM uncertainties. For example, at 8 TeV, if the ATLAS uncertainty is considered,
the aforementioned ratio results 7.5 standard deviation.
Besides the TOTEM results for $\sigma_{tot}$ being larger than the ATLAS values at 7 and 8 TeV, the
TOTEM data indicate
a rise of the total cross section faster than the ATLAS data \cite{fms17a}.

Obviously, these facts make any amplitude analyses more difficult and put
serious limitations in  secure interpretations of the results and  unquestionable conclusions
that may be reached. It is expected that these discrepancies might be resolved
through further re-analyses and/or new data, but it can also happen that these
systematic differences may persist. We recall the discrepancies characterizing the experimental
information at the highest energy reached in $\bar{p}p$ scattering, namely
1.8 TeV. The CDF and E710 results differ by 2.3 standard deviation (respect
the E710 uncertainty) and predictions from most phenomenological models
lies between these points.

Anyway, presently we understand 
that ensemble T+A is the effective representative of the experimental information
available, so that an efficient model should be able to access
all points within the corresponding uncertainties. More precisely, the predicted uncertainty region must present
agreement with the error bars of the experimental points, by reaching all
of them, even if in a barely way, but never excluding one or another data, namely TOTEM or ATLAS results.

On the basis of these comments and before discussing the efficiency of each model in the fit results,
three characteristics of ensembles T and T+A in our data reductions deserve to be highlighted.

\begin{itemize}

\item
From the Figures, for all models the main visual difference in the results
within ensembles T and T+A concerns $\sigma_{tot}$ at the highest energies but
not $\rho$ at these energies. Indeed, for example, with model I
(Fig. 1) the uncertainty region in the fit result for $\sigma_{tot}$ at 13 TeV
within ensemble T goes through the lower error bar, the central value and half
of the upper error bar, but within ensemble T+A, goes through only the lower
error bar; on the other hand, for $\rho$ at 13 TeV the uncertainty regions 
within T and T+A are essentially 
the same, lying far above the experimental data and error bars.
Analogous behaviors can be seen in Figs. 2, 3 and 4.
This is a consequence of the large number of experimental data on $\sigma_{tot}$
at the highest energies (mainly LHC region) as compared with those respect to $\rho$.

\item
From Tables I and II, in all cases (independently of the ensemble or model), 
for $\nu \sim$ 250, the $\chi^2/\nu$ lies in the region $\sim$ 1.2 - 1.3
and the integrated probability  $P(\chi^2) \sim 10^{-2} - 10^{-3}$.
Taking into account the discrepant values between TOTEM and ATLAS data, the
fits can be considered as reasonably accurate.

\item
For models
I, II and III the integrated probability $P(\chi^2)$ is one order of magnitude
smaller within ensemble T+A than within T and for model IV nearly 1/2.
This is a consequence of the aforementioned tension between the TOTEM and
ATLAS data at 7 and mainly 8 TeV.

\end{itemize}

\subsection{Models}

First, notice that from the Figures and within the uncertainties, all models present quite
good descriptions of the experimental data up to 7 TeV, as expected. Therefore, let us focus the discussion
in the region 8 - 13 TeV (mainly 13 TeV) and in the goodness of the fits.

\begin{enumerate}

\item Model I (DL-type)

The fit result in Fig. 1 is in plenty agreement with the $\sigma_{tot}$ datum at 13 TeV
within ensemble T and the uncertainty region crosses the lower error bar in case of ensemble T+A.
However, for $\rho$ the curves do not decrease in the region $10^3 - 10^4$ GeV (see insets)
and even with 2 $\sigma$ the results at 13 TeV lie
far above the upper error bars.
Within both ensembles the integrated probability is the smallest among the models ($10^{-3} - 10^{-4}$).
We conclude that this model is not  in agreement with the TOTEM data at 13 TeV.

\item Model II (COMPETE-type)

From Fig. 2 and ensemble T, the fit result (uncertainty region) for $\sigma_{tot}$ at 13 TeV crosses the 
central value and the lower
error bar and reaches half this bar within ensemble
T+A. For $\rho$, the curves decrease in the region $10^3 - 10^4$ GeV, but
as in the previous case, the uncertainty regions lie far above the
upper error bars (insets). We conclude this model does not present a satisfactory
description of the new data at 13 TeV.

\item Model III (BH-type)

From the Tables, the integrated probability is one of the highest among the models. From Fig. 3, 
for $\sigma_{tot}$ and ensemble T+A, the uncertainty region with 1 $\sigma$ reaches the upper error bar of the
ATLAS datum at 8 TeV and the lower bar  of the TOTEM datum at 13 TeV (similar with 2 $\sigma$ in case
of ensemble T). For $\rho$ the curves present the faster decrease among the models in the
region $10^3 - 10^4$ GeV (insets) and at 13 TeV, with 2 $\sigma$, the uncertainty region reaches the upper
extremum of the error bar with ensemble T+A (barely reach this point with ensemble T).
We understand that this model is not excluded by the bulk of experimental data presently available.

\item Model IV (Hybrid power-log)

Based on the disagreement of Model I with the TOTEM data at 13 TeV and given the efficiency
of the power law (simple pole Pomeron) below 13 TeV, we have tested hybrid contributions
by adding either a double pole or triple pole contributions. In the latter case the fits
did not converge and in the former case the fit results are presented in Tables I and II
and Figure 4. In this case we have one more parameter (as compared with 7 parameters in the
other 3 models), resulting in lager uncertainty regions. For $\sigma_{tot}$ the uncertainty
regions with 1 $\sigma$ encompass all the experimental data at the LHC energy region.
However, from the Tables the integrated probabilities are the smallest among the models  and 
although the results for $\rho$ (Figure 4) present a small decrease in the region
$10^3 - 10^4$ GeV, the uncertainty regions lie far above the TOTEM data.
We conclude that the model does not present agreement with the TOTEM data at 13 TeV. 

\end{enumerate}

\subsection{Conclusions on the Pomeron Models and Further Tests}

Based on the above discussion, we understand that models I, II and IV are not able to
describe simultaneously the TOTEM data on $\sigma_{tot}$ and $\rho$ at 13 TeV.
On the other hand, taking into account the bulk of experimental data presently available
(ensemble T+A) and the uncertainties in both theoretical and experimental results,
model III seems not to be excluded.

Looking for possible improvements in the efficiency of Model III,
we have also developed further tests with some variants.
Here, in all fits we have considered the energy cutoff at $\sqrt{s}_{\mathrm{min}}$ = 5 GeV
and the subtraction constant as a free fit parameter.
In order to investigate
the effect of the energy cutoff and the role of the subtraction constant  we have also carried out
fits without this parameter, namely by fixing $K = 0$ and rising the energy cutoff to
7.5 and 10 GeV. The results are presented in the Appendix, together with a short discussion.
Taking into account the energy region analyzed, 5 GeV - 13 TeV and the $pp$ and $\bar{p}p$
scattering, we did not find remarkable or considerable improvements. Indeed, with the cutoff at
5 GeV, the results with and without the subtraction constant are similar, with integrated probability
slightly greater in case of $K$ free and the uncertainty region reaching the extreme
of the upper error bar of $\rho$ at 13 TeV (ensemble T+A).

Therefore, we select as our best result those obtained here with model III, cutoff at 5 GeV and
the subtraction constant as a free fit parameter (Fig. 3 and Tables I and II).
For this case we present in Figure 5 a detail of the predictions for $\sigma_{tot}$ and $\rho$ at 13 TeV
and the experimental data; the numerical values are given in Table III, together with
the corresponding predictions at 14 TeV and uncertainties associated with 1 $\sigma$ and also 2 $\sigma$.

\begin{figure}[!ht]
\begin{center}
 \includegraphics[width=11.0cm,height=10cm]{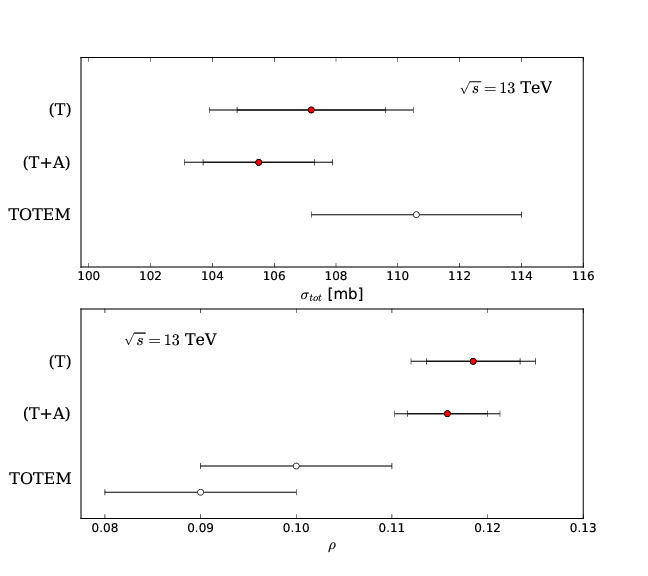}
 \caption{Predictions of Model III for $\sigma_{tot}$ and $\rho$ at $13$ TeV with $1$ and $2$ standard deviations from fits to ensemble T and T+A (filled circles) together with the TOTEM measurements $(3)$ (empty circles).}
\label{f5}
\end{center}
\end{figure}

\begin{table*}[!ht]
\caption{Predictions of Model III for $\sigma_{tot}$ and $\rho$ at 13 TeV and 14 TeV for $pp$ and $\bar{p}p$ 
scattering: central values and uncertainties 
with 1 $\sigma$ and 2 $\sigma$ (Tables I and II).}
\scalebox{0.8}{
\begin{tabular}{c@{\quad}c@{\quad}c@{\quad}c@{\quad}c@{\quad}c@{\quad}c@{\quad}c@{\quad}c@{\quad}}
\hline \hline
& & & & & & & &  \\[-0.2cm]
& & \multicolumn{3}{c} {$\sigma_{tot}$ (mb)} & & \multicolumn{3}{c}
{$\rho$} \\
\cline{3-5} \cline{7-9}
& & & & & & & &  \\[-0.2cm]
$\sqrt{s}$ (TeV) & Ensemble    & Central & $1\sigma$ & $2\sigma$ & & Central &
$1\sigma$ & $2\sigma$ \\[0.05ex]
\hline
& & & & & & & & \\[-0.3cm]
\multirow{2}{*}{13} & T     & 107.2 &  $\pm$\,2.4 & $\pm$\,3.3    & & 0.1185 &
$\pm$\,0.0049   & $\pm$\,0.0065 \\[0.05ex]
                    & T+A   & 105.5 &  $\pm$\,1.8 & $\pm$\,2.4    & & 0.1158 &
$\pm$\,0.0042   & $\pm$\,0.0055 \\[0.05ex]
\hline 
& & & & & & & & \\[-0.3cm]
\multirow{2}{*}{14} & T     & 108.4 & $\pm$\,2.5 & $\pm$\,3.3     & & 0.1179 &
$\pm$\,0.0049   & $\pm$\,0.0065 \\[0.05ex]
                    & T+A   & 106.7 & $\pm$\,1.8 & $\pm$\,2.5     & & 0.1152 &
$\pm$\,0.0042   & $\pm$\,0.0055 \\[0.05ex]
\hline \hline 
\end{tabular}
}
\label{t3}
\end{table*}

\newpage

\section{Conclusions and Final Remarks}

We have presented a forward amplitude analysis on the experimental data presently
available from $pp$ and $\bar{p}p$ scattering in the energy region
5 GeV - 13 TeV.
The analysis consists of tests with different analytic parameterizations for $\sigma_{tot}(s)$
and $\rho(s)$, all of them characterized by Pomeron leading contributions 
(even-under-crossing). 
The data reductions show that most models present no simultaneous agreement with the
recent $\sigma_{tot}$
and $\rho$ measurements at 13 TeV by the TOTEM Collaboration.
Different models and variants have been tested and among them,
Model III (two simple poles Reggeons, one double pole and one triple pole Pomerons), with 
only seven free fit parameters, led to the best results.

Two aspects have been stressed along the paper. The first concerns the TOTEM results
at 13 TeV, indicating an expected rise of the total cross section but an unexpected
decrease in the value of the $\rho$ parameter. The extrapolation from the recent analysis
with data up to 8 TeV,  discussed in our introduction,
shows clearly the plenty agreement with the $\sigma_{tot}$ result and the overestimation of
the $\rho$ data \cite{fms17a}. Note also that the value here obtained for the Pomeron
intercept with Model I and Ensemble T, $\epsilon$ = 0.0914 $\pm$ 0.0039 (Table I) is consistent
with results of fits up to 8 TeV, for example those obtained in \cite{ms13b}:
$\epsilon$ = 0.0926 $\pm$ 0.0016. However, the Model I result for $\rho$ at 13 TeV is
in complete disagreement with the TOTEM data (Figure 1). 

The second aspect
concerns the tension between the TOTEM and ATLAS data at 7 TeV and mainly at 8 TeV,
discussed in certain detail in the previous sections. That led us to consider separately
the two ensembles denoted T (excluding the ATLAS data) and T+A (including the ATLAS data).
We have shown that these discrepancies play an important role in the interpretations
of the fit results.

Another aspect deserves attention when interpreting the data reductions. As discussed
in Appendix A of \cite{fms17b}, the TOTEM uncertainties are essentially systematic (uniform distribution)
and not statistical (Gaussian distribution). Therefore, a model result crossing the central
value of an experimental result may have a limited significance
on statistical grounds.

The unexpected decrease in the $\rho$ value has been well described in the recent analyses
by Martynov and Nicolescu. The first paper treated only the TOTEM data \cite{martynico1} and 
in the second one the ATLAS data have been included \cite{martynico2}.
The $\chi^2/\nu$ are similar in both cases, namely 1.075 without ATLAS and 1.100 including ATLAS,
corresponding to an increase of 2.3 \%.
For $\rho$ at 13 TeV, in both cases the curves seems to cross the central value of
the experimental points (one symbol in Figures 2 and 3 of \cite{martynico2}). However, for $\sigma_{tot}$ with ATLAS excluded
the curve crosses the lower error bar at 13 TeV, but lies above the error bars of
the ATLAS data at 7 TeV and mainly 8 TeV. With ATLAS included, the curve crosses the
ATLAS data, but lies below the lower error bar of the TOTEM data at 8 TeV and
mainly 13 TeV (Figure 3 of \cite{martynico2}) . Summarizing, the curve does not reach
the upper error bars of the ATLAS data on $\sigma_{tot}$ at 7 and 8 TeV in the former case and does not reach
the lower error bar of the TOTEM datum at 13 TeV in the latter case.

In what concerns our results with Model III, the $\chi^2/\nu$ are also similar in both cases:
1.210 (T) and 1.234 (T+A), corresponding to an increase slightly small, 2.0 \%.
The uncertainty regions of the fit results do not cross the central
values of the $\rho$ data at 13 TeV, but barely reach the upper error bar. However, the same is
true for the ATLAS datum on $\sigma_{tot}$ at 8 TeV. Therefore, we conclude that the agreement
between the phenomenological model and the experimental points is reasonably compatible
within the uncertainties. In other words,
in case of fits to
ensembles T or T + A (all the experimental data presently available) 
and within the uncertainties, the Pomeron model III, with 7 free fit parameters, seems not to be excluded by the
experimental data presently available on forward $pp$ and $\bar{p}p$ elastic scattering.

In the theoretical context, the Odderon is a well-founded concept in
perturbative QCD \cite{braun}.
Despite the consistent phenomenological description of the unexpected
decrease of the $\rho$ parameter at 13 TeV, the Odderon
model predicts a crossing in the $pp$ and $\bar{p}p$ total cross sections
at high energies.
Although in agreement with high-energy theorems \cite{cross}, it seems
still lacking a pure
(model independent) nonperturbative QCD explanation (from first
principles) for an asymptotic rise of the
total cross section faster for hadron-hadron than for antihadron-hadron
collisions.

Finally, we understand that further re-analysis and new experimental data at 13 TeV and 14 TeV,
by the TOTEM and ATLAS collaborations, shall be crucial for
confronting, in a conclusive way, the possible dominance of Odderon or Pomeron 
in forward elastic hadron scattering at high energies.

\acknowledgments

This research was partially supported by the Conselho Nacional de
Desenvolvimento Cient\'{\i}fico e Tecnol\'ogico (CNPq). EGSL acknowledges
the financial support from the Rede Nacional de Altas Energias (RENAFAE).

\appendix*

\section{Further Tests with Model III}

In Sect. III the fits through model III to ensembles T and T+A were carried out with
energy cutoff at 5 GeV and the subtraction constant as a free fit parameter
(Tables I and II and Figure 3). 
In what follows, we consider two variants related to the energy cutoff and the subtraction constant.
Firstly, still with the subtraction constant as a free fit parameter, we develop fits with
energy cutoff at 7.5 and 10 GeV. The results are displayed in Table IV and Figures 6 and 7.
In a second step the subtraction constant is fixed at zero
and the fits are developed with energy cutoff at 5, 7.5 and 10 GeV.
The results are shown in Table V and Figures 8, 9 and 10.
As before, in all the cases we employ ensembles T and T+A and CL with one and two
standard deviations.

\begin{table*}[!ht]
\caption{Fit results with model III to ensembles T and T+A by considering one-standard deviation, energy cutoffs 
at 7.5 and 10 GeV and $K$ as a free fit parameter.}
\begin{tabular}{c@{\quad}c@{\quad}c@{\quad}c@{\quad}c@{\quad}c@{\quad}c@{\quad}c@{\quad}}
\hline \hline
& & & &    \\[-0.2cm]
Ensemble& \multicolumn{2}{c} {T} & & \multicolumn{2}{c} {T+A} \\
\cline{2-3} \cline{5-6}
& & & & & &  \\[-0.2cm]
$\sqrt{s}_{\mathrm{min}}$ (GeV)    & 7.5 & 10 &  & 7.5 & 10 \\[0.05ex]
\hline
& & & & & & \\[-0.2cm]
$a_{1}$ (mb)       & 57.5\,$\pm$\,2.1    & 55.8\,$\pm$\,4.0    &  & 57.9\,$\pm$\,2.1    & 56.5\,$\pm$\,4.1 \\[0.05ex]
$b_{1}$            & 0.217\,$\pm$\,0.023 & 0.202\,$\pm$\,0.037 &  & 0.224\,$\pm$\,0.021 & 0.212\,$\pm$\,0.037 \\[0.05ex]
$a_{2}$ (mb)       & 16.8\,$\pm$\,2.7    & 15.1\,$\pm$\,4.6    &  & 16.8\,$\pm$\,2.7    & 15.1\,$\pm$\,4.8 \\[0.05ex]
$b_{2}$            & 0.542\,$\pm$\,0.046 & 0.520\,$\pm$\,0.070 &  & 0.542\,$\pm$\,0.046 & 0.520\,$\pm$\,0.072 \\[0.05ex]
$C$ (mb)           & 3.48\,$\pm$\,0.44   & 3.25\,$\pm$\,0.66   &  & 3.66\,$\pm$\,0.38   & 3.48\,$\pm$\,0.59 \\[0.05ex]
$D$ (mb)           & 0.143\,$\pm$\,0.030 & 0.156\,$\pm$\,0.040 &  & 0.128\,$\pm$\,0.024 & 0.138\,$\pm$\,0.035 \\[0.05ex]
$K$ (mbGeV$^{2}$)  & -15\,$\pm$\,74      & 4.17\,$\pm$\,116    &  & -9.5\,$\pm$\,73     & 14.3\,$\pm$\,117 \\[0.05ex]
\hline
& & & & & & \\[-0.4cm]
$\nu$              & 205                    & 164                    &  & 207                    & 166 \\[0.05ex]
$\chi^2/\nu$       & 1.217                  & 1.213                  &  & 1.253                  & 1.263 \\[0.05ex]
$P(\chi^2)$        & 1.8 $\times$ 10$^{-2}$ & 3.3 $\times$ 10$^{-2}$ &  & 7.8 $\times$ 10$^{-3}$ & 1.2 $\times$ 10$^{-2}$ \\[0.05ex]
\hline
Figure             &          6             &           7            &  &    6                   &  7        \\[0.05ex]
\hline \hline 
\end{tabular}
\label{t4}
\end{table*}

\begin{table*}[!ht]
\caption{Fit results with model III to ensembles T and T+A by considering one-standard deviation,
energy cutoffs at 5, 7.5 and 10 GeV  and the subtraction constant $K=0$ (fixed).}
\scalebox{0.85}{
\begin{tabular}{c@{\quad}c@{\quad}c@{\quad}c@{\quad}c@{\quad}c@{\quad}c@{\quad}c@{\quad}}
\hline \hline
& & & & & & \\[-0.4cm]
Ensemble & \multicolumn{3}{c} {T} & & \multicolumn{3}{c} {T+A} \\
\cline{2-4} \cline{6-8}
& & & & & &  \\[-0.2cm]
$\sqrt{s}_{\mathrm{min}}$ (GeV)    & 5 & 7.5 & 10 & & 5 & 7.5 & 10 \\[0.05ex]
\hline
& & & & & & \\[-0.2cm]
$a_{1}$ (mb)      & 58.6\,$\pm$\,1.3    & 57.7\,$\pm$\,1.8    & 55.8\,$\pm$\,3.1    & & 58.8\,$\pm$\,1.3    & 58.0\,$\pm$\,1.8    & 56.2\,$\pm$\,3.1 \\[0.05ex]
$b_{1}$           & 0.226\,$\pm$\,0.015 & 0.219\,$\pm$\,0.019 & 0.202\,$\pm$\,0.029 & & 0.231\,$\pm$\,0.014 & 0.225\,$\pm$\,0.018 & 0.209\,$\pm$\,0.028 \\[0.05ex]
$a_{2}$ (mb)      & 17.0\,$\pm$\,1.8    & 16.6\,$\pm$\,2.3    & 15.2\,$\pm$\,4.2    & & 17.1\,$\pm$\,1.8    & 16.6\,$\pm$\,2.3    & 15.3\,$\pm$\,4.3 \\[0.05ex]
$b_{2}$           & 0.547\,$\pm$\,0.032 & 0.538\,$\pm$\,0.038 & 0.521\,$\pm$\,0.064 & & 0.548\,$\pm$\,0.032 & 0.539\,$\pm$\,0.038 & 0.522\,$\pm$\,0.063 \\[0.05ex]
$C$ (mb)          & 3.62\,$\pm$\,0.30   & 3.51\,$\pm$\,0.38   & 3.24\,$\pm$\,0.53   & & 3.76\,$\pm$\,0.26   & 3.67\,$\pm$\,0.33   & 3.44\,$\pm$\,0.47 \\[0.05ex]
$D$ (mb)          & 0.135\,$\pm$\,0.022 & 0.141\,$\pm$\,0.026 & 0.157\,$\pm$\,0.033 & & 0.122\,$\pm$\,0.018 & 0.127\,$\pm$\,0.021 & 0.140\,$\pm$\,0.029 \\[0.05ex]
\hline
& & & & & & \\[-0.4cm]
$\nu$             & 249       & 206       & 165                 & & 251                 & 208                 & 167 \\[0.05ex]
$\chi^2/\nu$      & 1.210     & 1.213     & 1.206               & & 1.238               & 1.248               & 1.256 \\[0.05ex]
$P(\chi^2)$       & 1.3 $\times$ 10$^{-2}$ & 2.0 $\times$ 10$^{-2}$ & 3.7 $\times$ 10$^{-2}$ & & 6.1 $\times$ 10$^{-3}$ & 8.8 $\times$ 10$^{-3}$ & 1.4 $\times$ 10$^{-2}$ \\[0.05ex]
\hline
Figure            &    8      &      9    &         10             & & 8           &     9   &  10        \\[0.05ex]
\hline \hline 
\end{tabular}
}
\label{t5}
\end{table*}

\begin{figure}[H]
\begin{center}
 \includegraphics[width=8.0cm,height=9.5cm]{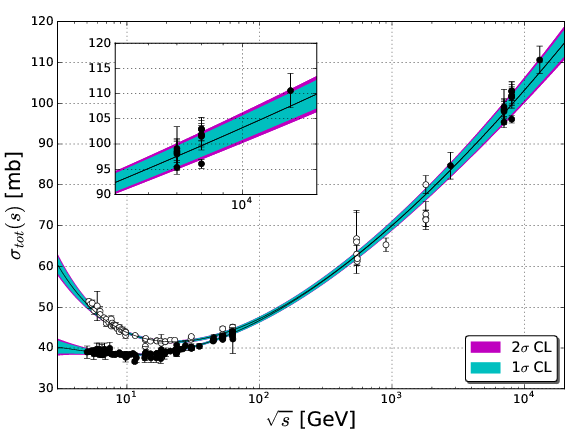}
 \includegraphics[width=8.0cm,height=9.5cm]{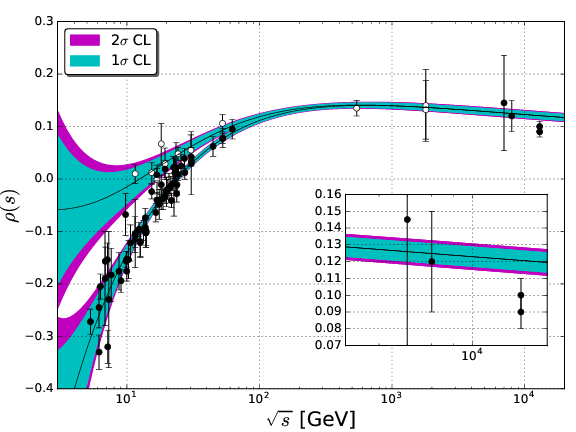}
 \includegraphics[width=8.0cm,height=9.5cm]{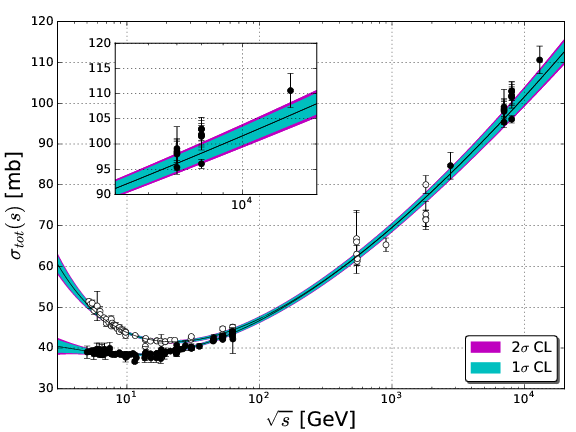}
 \includegraphics[width=8.0cm,height=9.5cm]{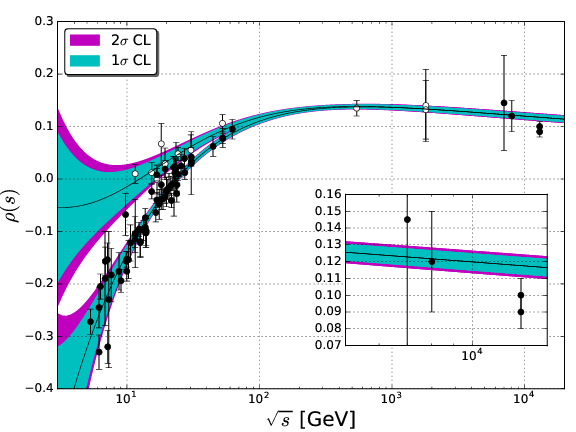}
 \caption{Fit results with Model III to ensembles T (above) and T+A (below) by considering the 
energy cutoff at $\sqrt{s}=7.5$ GeV and $K$ as a free fit parameter.}
\label{f6}
\end{center}
\end{figure}

\begin{figure}[H]
\begin{center}
 \includegraphics[width=8.0cm,height=9.5cm]{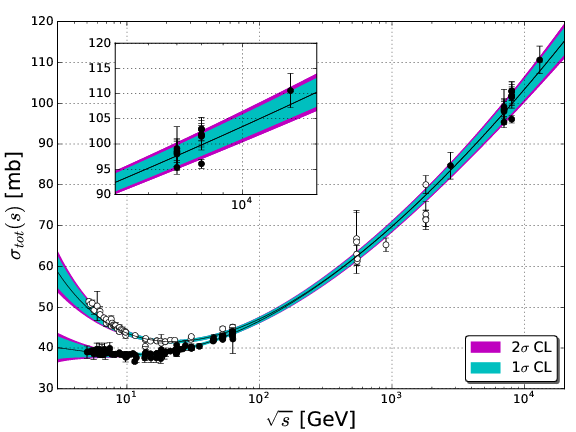}
 \includegraphics[width=8.0cm,height=9.5cm]{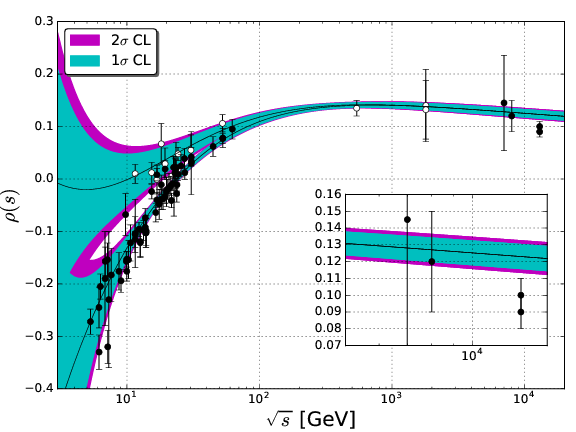}
 \includegraphics[width=8.0cm,height=9.5cm]{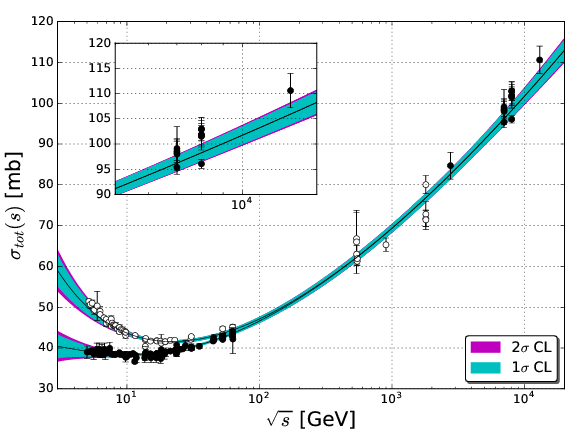}
 \includegraphics[width=8.0cm,height=9.5cm]{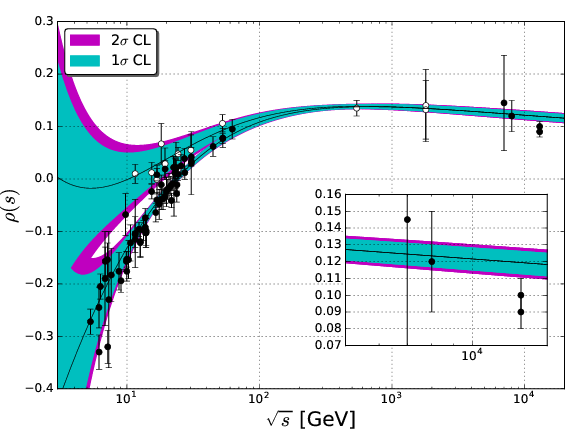}
 \caption{Fit results with Model III to ensembles T (above) and T+A (below) by considering 
the energy cutoff at $\sqrt{s}=10$ GeV and $K$ as a free fit parameter.}
\label{f7}
\end{center}
\end{figure}

\begin{figure}[H]
\begin{center}
 \includegraphics[width=8.0cm,height=9.5cm]{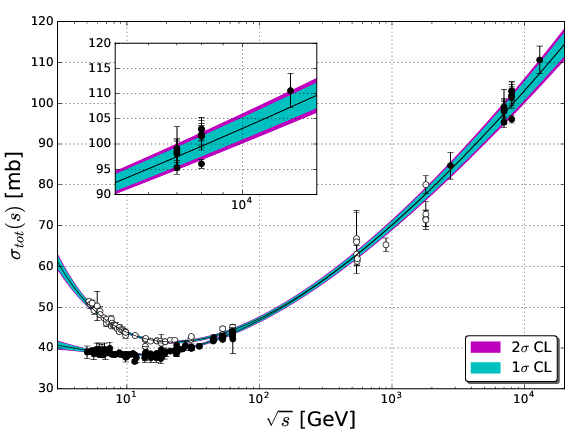}
 \includegraphics[width=8.0cm,height=9.5cm]{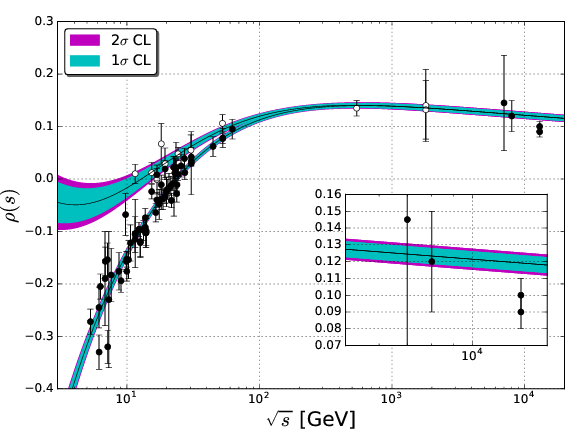}
 \includegraphics[width=8.0cm,height=9.5cm]{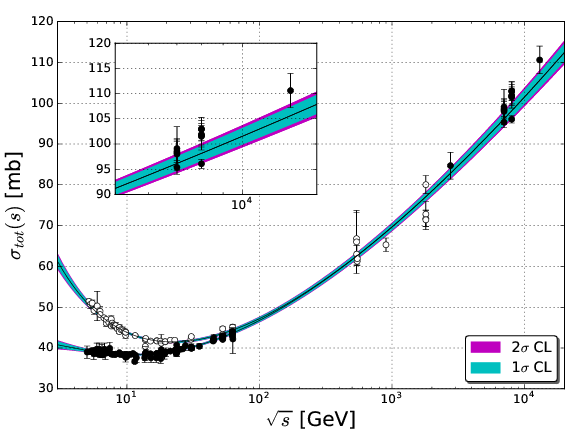}
 \includegraphics[width=8.0cm,height=9.5cm]{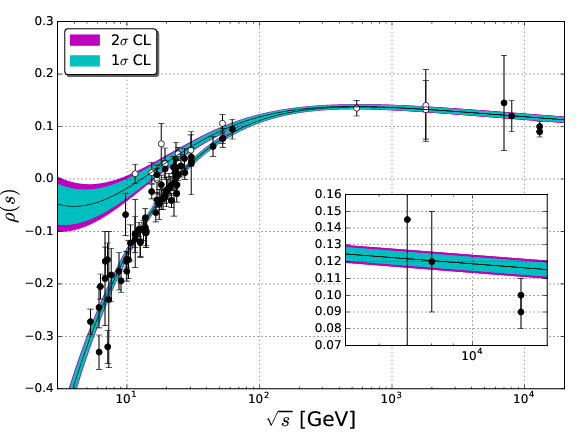}
 \caption{Fit results with Model III to ensembles T (above) and T+A (below) by considering 
the energy cutoff at $\sqrt{s}=5$ GeV and $K=0$ (fixed).}
\label{f8}
\end{center}
\end{figure}

\begin{figure}[H]
\begin{center}
 \includegraphics[width=8.0cm,height=9.5cm]{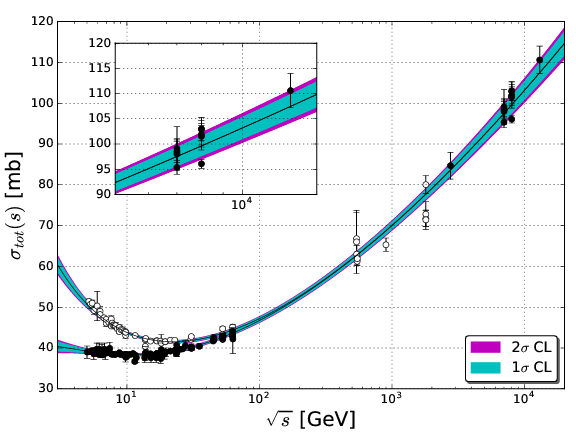}
 \includegraphics[width=8.0cm,height=9.5cm]{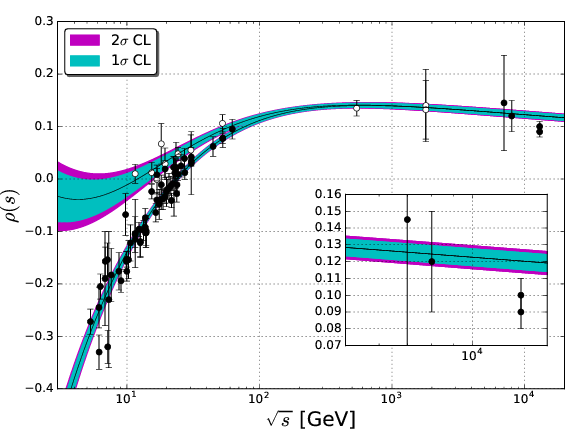}
 \includegraphics[width=8.0cm,height=9.5cm]{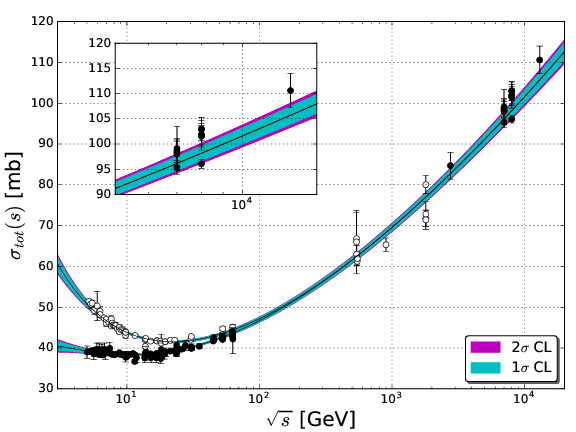}
 \includegraphics[width=8.0cm,height=9.5cm]{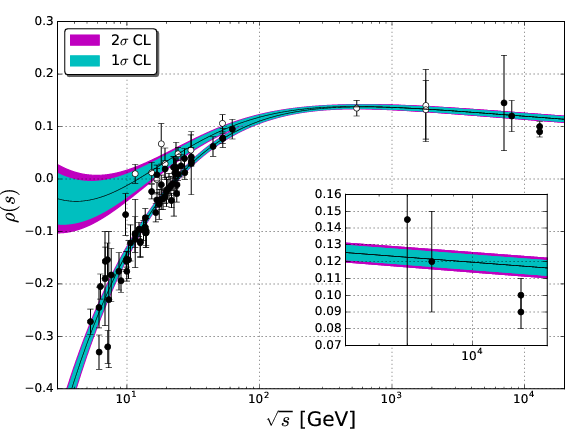}
 \caption{Fit results with Model III to ensembles T (above) and T+A (below) by considering 
the energy cutoff at $\sqrt{s}=7.5$ GeV and $K=0$ (fixed).}
\label{f9}
\end{center}
\end{figure}

\begin{figure}[H]
\begin{center}
 \includegraphics[width=8.0cm,height=9.5cm]{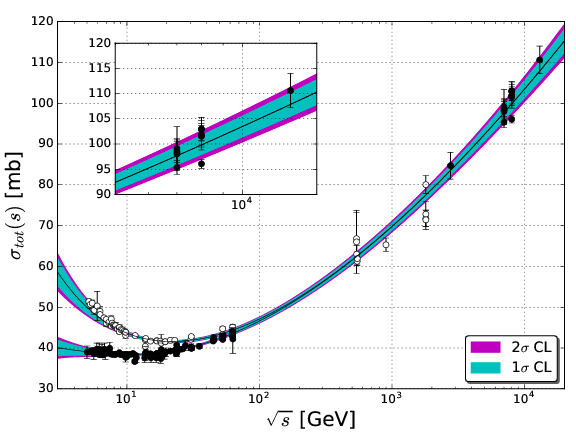}
 \includegraphics[width=8.0cm,height=9.5cm]{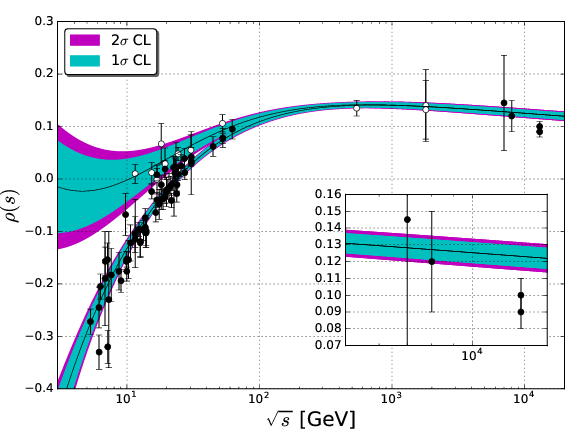}
 \includegraphics[width=8.0cm,height=9.5cm]{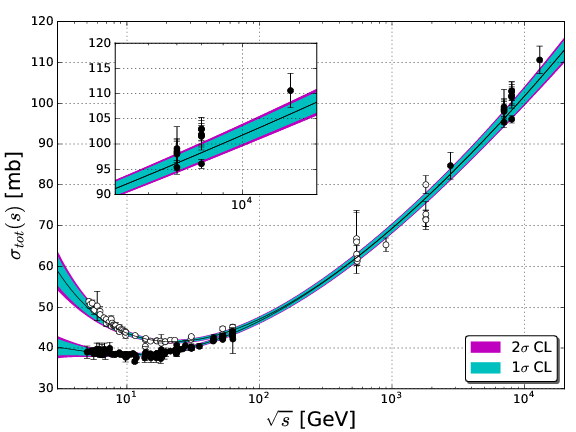}
 \includegraphics[width=8.0cm,height=9.5cm]{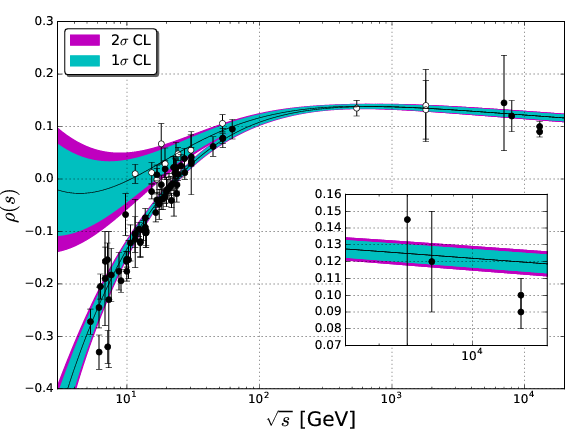}
 \caption{Fit results with Model III to ensembles T (above) and T+A (below) by considering the 
energy cutoff at $\sqrt{s}=10$ GeV and $K=0$ (fixed).}
\label{f10}
\end{center}
\end{figure}

\newpage

For K as a free fit parameter, comparison of Table I (cutoff at 5 GeV) with Table IV
(cutoffs at 7.5 and 10 GeV), shows that for both ensembles, rising the
cutoff results in a slightly increase in $P(\chi^2)$ and from Figures
3, 6 and 7, the uncertainty regions become larger, mainly at lower energies.
The same effect is observed by fixing $K = 0$  (Table V and Figures
8, 9 and 10). The rise of the cutoff does not led to an improvement
in the fit results, within the uncertainty region, at 13 TeV.

For cutoff at 5 GeV, the results with $K$ free (Table I, Figure 3)
and $K=0$ fixed (Table V, Figure 8) show the following features:

\begin{itemize}

\item
within both ensembles, the integrated probability is slightly larger for $K$ free;

\item
for $\rho$ at 13 TeV and ensemble T, the distance between the minimum of
the uncertainty region and the extreme of the upper error bar is smaller with
$K$ free than with $K=0$;

\item
for $\rho$ at 13 TeV, ensemble T+A and $K=0$, the uncertainty region lies slightly
above the extreme of the upper error bar (Figure 8) and for $K$ free the uncertainty
region reaches this point (Figure 3).

\end{itemize}

In conclusion, the rising of the cutoff does not lead to improvements in the fit results,
neither fixing $K=0$. The results with $K$ free and cutoff at 5 GeV present best 
agreement with the TOTEM data at 13 TeV.

\end{document}